\documentclass[onecolumn,12pt]{IEEETran}
\usepackage{bbm}
\usepackage{amsfonts}
\usepackage{tipa}
\usepackage{amssymb}
\usepackage{mathrsfs}
\usepackage[lined,boxed,commentsnumbered]{algorithm2e}
\usepackage{graphicx}
\usepackage{amsfonts,amssymb,amsmath}
\usepackage{latexsym}
\usepackage{color, soul}
\usepackage{multirow}
\usepackage{hyperref}
\usepackage{epsfig,epstopdf}
\usepackage{array}

\newtheorem{thm}{Theorem}
\newtheorem{lemma}{Lemma}

\begin{document}
\title{
Compute-and-Forward: Optimization Over Multi-Source-Multi-Relay Networks}
\author{Zhi~Chen~\IEEEmembership{Member,~IEEE,}, Pingyi~Fan~\IEEEmembership{Senior Member,~IEEE,}
        and~Khaled~Ben~Letaief~\IEEEmembership{Fellow,~IEEE}
\thanks{Z. Chen and P. Fan are with the Department of Electrical Engineering, Tsinghua University, Beijing, China, 100084. Emails: chenzhi06@mails.tsinghua.edu.cn; fpy@tsinghua.edu.cn.

K. B. Letaief is with Department of Electronic and Computer Engineering,
Hong Kong University of Science and Technology, Hong Kong (e-mail:
eekhaled@ece.ust.hk).}}
\maketitle

\maketitle
\baselineskip 24pt
\begin{abstract}\\
\baselineskip=18pt
In this work, we investigate a multi-source multi-cast network with
the aid of {an arbitrary number of} relays, where it is assumed that no
direct link is available at each S-D pair. {The aim is
to find the fundamental limit on
the maximal common multicast throughput of all source nodes if resource allocations are available.} A transmission protocol employing
the relaying strategy, namely, compute-and-forward (CPF), is proposed.
{We also adjust the
methods in the literature to obtain the integer network-constructed
coefficient matrix (a naive method, a local optimal method as well as a global optimal method)
to fit for the general topology with an arbitrary number of relays.}
Two transmission scenarios are addressed. The first scenario is delay-stringent transmission where each message must be delivered within one slot.
The second scenario is delay-tolerant transmission where
no delay constraint is imposed. The associated optimization problems
to maximize the short-term and long-term common multicast throughput
are formulated and solved, and the optimal allocation of power and
time slots are presented. Performance comparisons show that the CPF
strategy outperforms conventional decode-and-forward (DF) strategy.
{It is also shown that with more relays, the CPF strategy performs even
better due to the increased diversity.}
Finally, by simulation, it is observed that for a large network in relatively
high SNR regime, CPF with the local optimal method for the
network-constructed matrix can perform close to that with the global
optimal method.
\baselineskip=18pt
\end{abstract}
\IEEEpeerreviewmaketitle

\section{Introduction}
Network coding, an efficient way to mitigate network interference and improve throughput, was firstly proposed by Yeung \emph{et al} in \cite{ahlswede2000network}\cite{li2003linear} for wireline networks. Employing it, relay nodes capability  is expanding from only simply forwarding messages to forwarding some functions of different messages to multiple destinations nodes. The intended nodes can then detract the message required as long as they have prior knowledge of the rest messages. In this way, wireline network throughput is improved. Furthermore, network coding is shown to be promising in wireless networks in terms of throughput improvement in \cite{fragouli2007wireless}\cite{li2010network}\cite{niu2010cooperative}\cite{zhang2009network}.

However, due to broadcast nature in wireless communications, the performance potential of conventional digital network coding (DNC) is strictly constrained by the interference from transmissions of other irrelevant transmitters in wireless networks. For instance, for a simple three-node, two-way relay network (TWRN), the relay node needs to jointly decode two individual messages in the multi-access phase with DNC \cite{oechtering2008stability}\cite{chen2012energy}\cite{Chen2012two-way5}, whereas the performance is degraded due to the fact that one user's message is regarded as interference to the message of the other user at the relay node in the multi-access uplink. As it is, this interference constrains the achievable rate pair especially in high signal-to-noise ratio (SNR) regime. Other strategies, like amplify-and-forward (AF) and compress-and-forward (CF), has its intrinsic advantages without decoding individual messages, but the noise term will be amplified along with messages delivery throughout the network, resulting in degradation of network performance.

To this end, a smart way of network coding, which was referred to as
physical layer network coding (PNC) \cite{popovski2007physical}
\cite{zhang2006physical}\cite{zhang2009channel}\cite{wang2009channel}
\cite{lu2011asynchronous}, i.e., compute-and-forward (CPF) \cite{nazer2011compute}
for general multi-source multi-relay networks, attracted increasing attention.
Typically for a TWRN, in the uplink phase, the two source nodes simultaneously
transmit their messages to the relay node, and the relay node merely decodes a
linear combination of these two messages with integer coefficients, other than
 employing joint decoding. In the downlink phase, the relay node can transmit to
 the two source nodes the linear combined  message. In this case, the two
 sources can subtract their transmitted messages and then decode the
 intended message. In this way, the relay node mitigates the
 interference coming from joint decoding in the uplink phase by
 employing digital network coding, and also avoids the noise amplification
 when using analog network coding (ANC) \cite{katti2007embracing}
 \cite{zhang2009optimal}. In \cite{narayanan2007joint}, PNC was shown
 to perform close to a capacity upper bound and its performance gain
 over DNC and ANC was demonstrated.

More importantly, PNC is shown to achieve high performance for more general
multi-source multi-relay networks in \cite{nazer2011compute}
\cite{nazer2011reliable}\cite{xiao2009design}\cite{xiao2010multiple}.
In the celebrated work \cite{nazer2011compute}, PNC was referred to
as compute-and-forward (CPF) strategy. With this strategy, each relay
will decode a function message formed by a linear combination of the
messages from all source nodes with a selected integer coefficient vector.
Each destination node hence obtains different function messages from
various relay nodes and decodes all the source messages as long as
the integer coefficients constructed matrix is in full rank. In the literature,
the outage probability performance of CPF is demonstrated to
outperform other relaying strategies, such as decode-and-forward (DF)
and amplify-and-forward (AF). In all these works,
a type of linear codes, lattice code were employed to achieve
the derived CPF capacity region.
On the other hand, Nazer \emph{et al} mainly investigated the outage rate
of the CPF rate over S-R links in \cite{nazer2011compute},
whereas the transmission over R-D links was not discussed.
In \cite{osmane2011compute}, how to obtain the locally optimal integer
coefficient vector at each relay node to maximize its computation rate
was addressed,
however the full rank of the matrix constructed by all the integer
coefficient vectors of all relays was not guaranteed hence the
destination nodes may still not be able to decode all source messages.
In \cite{wei2012compute} it jointly optimized the integer coefficient
vector at each relay to finally obtain the optimal common computation
rate with the guaranteed full rank matrix constructed by these integer
coefficient vectors, at the cost that the achievable computing rates at
some relay nodes may be reduced to satisfy the full rank requirement.
In addition, the outage performance of CPF under some specific network
configurations were addressed in the literature, e.g.,
\cite{osmane2011compute} for a three-transmitter multi-access
network and \cite{wang2012outage} for multi-way relay networks
with the aid of only one relay node.

{Note that all the previous works in the literature for general networks
only considered the outage performance of CPF strategy with
constant transmit power and no optimal resource allocation was
studied. In this work, we therefore aim to investigate the achievable throughput with CPF
with adjustable resource allocation strategies, assuming that
full channel state information (CSI) is available
at all transmitters. By doing so,
the fundamental limit on the maximal common multicast
throughput can be obtained, which is useful in
the system design of multi-source multi-relay networks.}

There are also some works investigating the potential
of lattice codes as a capacity-achieving codes in
\cite{narayanan2007joint}\cite{zamir2002nested}\cite{erez2004achieving}
\cite{gunduz2009multi}, where in \cite{zamir2002nested}
Zamir \emph{et al} discussed nested lattice codes for
structured multi-terminal binning, and in \cite{erez2004achieving}
Zamir \emph{et al} showed the capacity-achievable property of
lattice codes over AWGN channels. In \cite{gunduz2009multi},
lattice codes was employed for a multi-way relay network and
the capacity region to within a half-bit of the cut-set bound
was shown to be achievable with it.

In this work, a general multi-source multi-relay multicast network is considered and the lattice codes are employed to realize CPF transmission.
{The performance with CPF in terms of the fundamental limit on the achievable common
multicast throughput is investigated},
with both the S-R links as well as the R-D links taken into account.
Two cases of interest will be studied. One is a delay-stringent scenario,
where each multicast transmission from all sources to all destinations
must be finished in one slot. The other is a delay-tolerant
transmission scenario, where the multicast transmission from all sources
to all destinations can be finished within arbitrary finite number of slots, i.e., no delay constraints are imposed. The major contributions of this work
are listed as follows.
%
    \begin{itemize}
    \item We design a CPF based multi-source multicast transmission protocol
    for the topology {with an arbitrary number of relays}.
    \item
    For the delay-stringent scenario, an optimization problem to maximize the common multicast throughput over one block with the specified channel gains employing CPF, is formulated and solved.
    \item
    For the delay-tolerant scenario, an optimization problem to
    maximize the average common multicast throughput employing CPF
    by allocating time and power resources, is formulated and solved analytically.
    {In addition, the convexity of the formulated problem is
    proved analytically.}
    \item
    We find that through simulations,
    \begin{enumerate}
    \item with an arbitrary number of source nodes,
    CPF with global-optimized network-constructed matrix
    outperforms DF strategy, which verifies the superiority of CPF over DF.
    \item CPF performs better with the increasing number of relay nodes.
    \item with a small number of source nodes, CPF with local-optimized network-constructed matrix performs slightly worse than DF for delay-stringent scenario, while slightly better than DF for delay-tolerant scenario, due to higher rank failure probability.
    \item with a relatively large number of source nodes, CPF with local
    optimized network-constructed matrix approximates
    the performance of CPF with global optimized network-constructed
    matrix and outperforms DF, due to the reduced rank failure probability.
    {This finding makes the implementation of CPF more practical and flexible,
    due to the greatly reduced network overhead
    information exchange required by the global optimized method
    for forming matrix.}
    \end{enumerate}
\end{itemize}

The remainder of this work is organized as follows.
In Section II, we present the system model of a multi-source
multicast network with the aid of multiple relay nodes,
and describes clearly the transmission protocol and the decoding procedure at the destination nodes.
In Section III and Section IV, delay-limited scenario and delay-tolerant
scenario are investigated, respectively.
{The associated problems to find the fundamental limit on the maximal common multicast throughput
of the entire network} are formulated and solved, by jointly allocating
time and energy resources for each transmit phase.
Simulation results are presented in Section V.
Finally, we conclude this work in Section VI.

\begin{figure}[!h]
   \centering
   \includegraphics[width = 6cm]{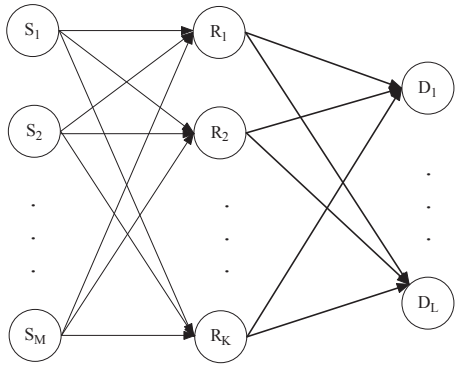}
   \caption{System model for a multi-source multicast network with the aid of multiple relay nodes. The direct link between each S-D pair is assumed to be unavailable.} \label{fig:system_model}
   \end{figure}

\section{System Model}
In this work, we mainly focus on a multi-source, multi-relay, multicast network, as shown in Fig. \ref{fig:system_model}, which consists of $M$ sources ($S_1$,\ldots, $S_M$), $K$ relays ($R_1$,\ldots, $R_K$) and $L$ destinations ($D_1$,\ldots, $D_L$). Each source node or relay node is
equipped with one antenna and works in half-duplex mode, i.e., cannot
transmit and receive data simultaneously. No direct link between any
S-D pair is assumed to be available. Hence, transmission must be assisted
by relay nodes.
A block fading channel model is also assumed for each link. The stochastic and instantaneous channel gain of each link are assumed to be known at both the transmitter and the receivers, which can be realized via feedback.
{In this work, lattice coding is adopted, as it can not only achieve close-to capacity rate,
but also preserves the linear property, i.e., the linear combination of lattice
codes is also a lattice point, which is essential for CPF strategies.}

We define $h_{ml}$ as the channel fading coefficient of the link $S_l$-$R_m$ and ${\bf z}_m$ as the i.i.d. additive white Gaussian noise vector, i.e., ${\bf z}_m \sim {\it N}({\bf 0},{\bf I}_n)$. We also denote ${\bf h}_m=[h_{m1} \ldots h_{mM}]^T$ by the channel coefficient vector consisting of the links from all sources to $R_m$. Similarly, we denote ${\bf g}_m=[g_{m1} \ldots g_{mL}]^T$ by the channel coefficient vector consisting of the links from the $m$th relay to all destinations and $g_{m_{\min}}=\min_i |g_{mi}|^2 $ as the minimum channel gain of the links from the $m$th relay to all destinations.
We also assume that all nodes are with the same average power constraint $P_0$ 
In addition, we assume the integer coefficient vector adopted at the $m$th relay in decoding the function message is ${\bf a}_m \in {\mathbb{Z}}^M$, which is carefully selected to form a
probably decodable integer-combined version. It is interestingly noted that
the $m$th relay can select different ${\bf a}_m$ to form different decodable function messages, albeit at different CPF rates.

\begin{figure}[!h]
   \centering
   \includegraphics[width = 7cm]{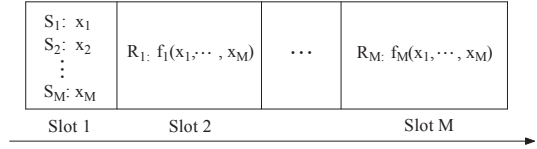}
   \caption{{Transmission protocol for a multi-source multicast network
   with the aid of relays. The direct link between each S-D pair is assumed
    to be unavailable. In this figure, each $S_i$ wants to broadcast a message
    $x_i$ to all destinations nodes and they simultaneously transmit
    in the first slot.
   In the following slots, each relay node forwards a decoded combined
   message
   to all destinations.
   For example, in the $i+1$th slot, $R_i$ broadcasts a message
   $f_i(x_1,x_2,\cdots,x_M)=\sum_{j=1}^M a_{ij}x_j$ to all destination nodes.}} \label{fig:transmission}
   \end{figure}

As shown in Fig. \ref{fig:transmission}, a CPF based transmission protocol consists of $M+1$ phases. The detailed procedure for the case $K=M$\footnote{It will be discussed in Section \ref{sec:general}
for the transmission procedure for the two cases of $K<M$ and $K>M$.} is illustrated as follows.
\begin{itemize}
\item In Phase $1$, all source nodes transmit their messages simultaneously to all the relay nodes. At the end of this phase, each relay node decodes one or more linear equations of the combination of individual transmitted messages from all sources with selected integer NC coefficient vectors.
\item In Phase $i$ ($i=2, \ldots, M+1$), the $(i-1)$th relay delivers its decoded function message to all destination nodes.
    At the end of Phase $i$, all the destination nodes decode the function message received and store it for source-message decoding at the end of Phase $M$.
\item Source-message decoding at the end of Phase $M+1$: with $M$ decoded function messages
from the relay nodes, each destination nodes tries to recover all original messages.
This would be possible if sufficient amount of equations are received reliably,
i.e., rank $M$ (the number of source nodes) of the matrix constructed by all these integer coefficients is achieved at each destination node.
\end{itemize}

Note that for the CPF phase (Phase 1), from \cite{nazer2011compute},
with the specified ${\bf a}_m$, the CPF rate at the $m$th relay
for the real-valued AWGN networks, achieved by
lattice coding, is
\begin{align}
R_{\mathrm{CPF}}^m = \max_{{\bf a}_m \in \mathbb{R}}\frac{1}{2}\log^+ \left( ||{\bf a}_m||^2 - \frac{P({\bf h}_m^{\mathrm{T}}{\bf a}_m)^2}{1+P||{\bf h}_m||^2} \right)^{-1}. \label{eqn:optrate}
\end{align}
where $P$ is the transmit power. The achievable common rate of all relays, is hence given by,
\begin{align}
R_{\mathrm{CPF}}=\min_i R^i_{\mathrm{CPF}} \quad i=1,\ldots, K.
\end{align}
Correspondingly, the required transmit power at all source nodes
for the $m$th relay with the common transmit rate $R_{\mathrm{CPF}}$, i.e., $P_{\mathrm{CPF}}^m$, is given by,
\begin{align}
P_{\mathrm{CPF}}^m=\frac{1-2^{2R_{\mathrm{CPF}}}b_m}{2^{2R_{\mathrm{CPF}}}c_m-a_m} \label{eqn:app1_2}
\end{align}
where
$a_m=|{\bf h}_m|^2$, $b_m=|{\bf a}_m|^2$,
$c_m=|{\bf h}_m|^2|{\bf a}_m|^2-|{\bf h}_m^{\mathrm{T}} {\bf a}_m|^2$ and
$d_m=|{\bf h}_m^{\mathrm{T}} {\bf a}_m|^2=a_mb_m-c_m$. 
The required transmit power at all source nodes for the CPF phase is given by,
$$P_{\mathrm{CPF}}=\max_m P_{\mathrm{CPF}}^m.$$ At this power level, all relays can enjoy the common computing rate $R_{\mathrm{CPF}}$.

In addition, for the achievable common CPF rate, we would like to show an interesting property of
the coefficients $a_m$, $b_m$ and $c_m$ related to $R_{\mathrm{CPF}}$,
which is summarized in Lemma \ref{appendix_1}.
\begin{lemma} \label{appendix_1}
With positive transmit power at relay nodes, we have
\begin{align}
\max_m (\frac{1}{b_m})<2^{2R_{\mathrm{CPF}}}<\min_m \frac{a_m}{c_m} \quad m=1,\ldots,K
\end{align}
\end{lemma}
The proof is given in Appendix \ref{appen_2} and is omitted here.

{On the other hand, for the relaying phases, the relay nodes forward the function messages
to all destination nodes consecutively. Since all destination nodes need to
successfully receive the function messages, the broadcast rate is determined by
the minimum channel gain of the R-D links for the transmitted relay node, i.e., $g_{m_{\min}}$. Hence, the
broadcast rate of the $i$th relay is determined by,}
 \begin{align}
 R_{r_i}=\frac{1}{2}\log_2(1+P_{r_i}g_{m_{\min}}).
 \end{align}

\subsection{Case Study: A Simple Example} \label{sec:case}
{Consider a three-source, three-relay and three-destination network.
The message transmitted by each source is $x_i$ ($i=1,2,3$) in the first hop.
The first relay decodes a function message of $2x_1+3x_2+5x_3$ ($z_1$) and forwards it to all destinations in the second hop.
The second and the third relay decode function messages of
$x_1+x_2+x_3$ ($z_2$) and $2x_1+x_2+5x_3$ ($z_3$) respectively. Hence they transmit
these function messages in the third hop and the fourth hop consecutively.
Assume that all destination nodes successfully receive the three function messages and attain the coefficient vectors of the three relays.
To obtain the three original source messages, all destination nodes
are then required to solve the following equation below.}
\begin{align}
\begin{bmatrix} 2 & 3 & 5 \\ 1 & 1 & 1\\ 2 & 1 & 5 \end{bmatrix}
\begin{bmatrix} x_1 \\ x_2\\ x_3 \end{bmatrix}
=A\begin{bmatrix} x_1 \\ x_2\\ x_3 \end{bmatrix}=\begin{bmatrix} z_1 \\ z_2\\ z_3 \end{bmatrix}
\end{align}
{To ensure the uniqueness of the solution to this matrix equation,
$A$ must be a full-rank matrix, i.e., $rank(A)=3$ for the given case
and each destination can then decode all the original messages from the sources.

Unfortunately, by letting $z_3=z_1=2x_1+3x_2+5x_3$,
we have $rank(A)=2$ and $x_1$ and $x_3$ can not be decoded at the destinations
and the entire transmission fails.
The importance of selection of the coefficient vectors to guarantee $A$ is
in full rank is therefore verified and we discuss it} in Sec. \ref{sec:vector} for $K=M$
and Sec. \ref{sec:general} for the general topology.

\begin{figure}[!t]
   \centering
   \includegraphics[width = 5.5cm]{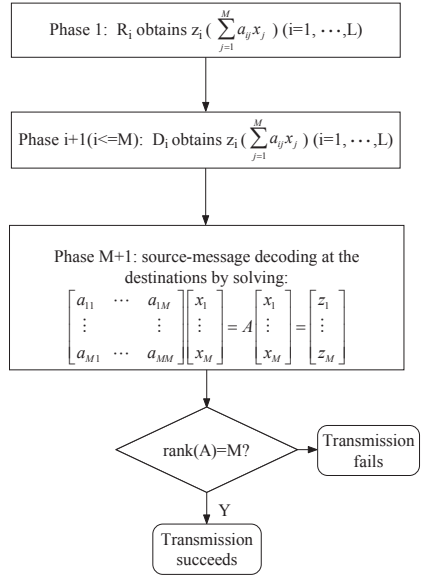}
   \caption{{Decoding procedure for a multi-source multicast network
   with the aid of relays. The direct link between each S-D pair is
   assumed to be unavailable. In this figure, $S_i$ wants to broadcast
   a message $x_i$ to all destinations nodes. They simultaneously transmit
   in the first slot and the relay node $R_i$ decodes a combined version
   of all source messages, $z_i=\sum_{i=1}^M a_{ij}x_j$.
   In each of the following slots, each destination
   node decodes a combined message from the relays and it
   tries to decode the source messages by solving a matrix
   equation at the end of the transmission procedure as depicted.}} \label{fig:decoding}
   \end{figure}

\subsection{Integer Coefficient Vector for the case $K=M$} \label{sec:vector}

From the discussion above, it is observed that selection of
coefficient vector is crucial for the achievable CPF rates.
Hence for clarity, we shall briefly review
the three methods in the literature to
obtain the integer coefficient vectors. {Note that all these
methods in the literature were only presented for the case $K=M$.}
For a given ${\bf h}_m$ at the $m$th relay, we can find ${\bf a}_m$ by
\begin{itemize}
\item method a) (naive method): obtain the integer coefficient vector ${\bf a}_m$ individually by solving $a_{mj}=\mathrm{round}(h_{mj})$ ($j=1,\ldots,M$) at the $m$th relay, where the function $\mathrm{round}(\cdot)$ returns the closest integer to $\{\cdot\}$.
\item method b) (local optimal method): obtain the locally-optimal integer coefficient vector ${\bf a}_m$ individually as in \cite{osmane2011compute} with the respective ${\bf h}_m$ at the $m$th relay.
\item method c) (global optimal method): obtain all the integer coefficient vectors (${\bf a}_m$) ($m=1,\ldots,M$) at all relays by employing the jointly optimization method in \cite{wei2012compute} to ensure that the full rank requirement of the matrix constructed by the integer coefficient vectors of all relays is satisfied.
\end{itemize}
For methods a) and b), only the local channel information is required. Therefore, they can not guarantee the full rank requirement at the destination nodes, i.e., the corresponding destination node may ultimately fail to decode the source messages and lead to the failure of the entire transmission.

On the other hand, method c) can guarantee the full matrix requirement
at the cost of indispensable additional signaling overhead among the relay nodes,
due to the joint optimal search procedure. It is therefore expected
to perform better than method a) and b) at the sacrifice of higher overhead.

\subsection{General Scenario} \label{sec:general}
{From the case study in} \ref{sec:case}, {for the general M-source multicast network,
we must also have $rank(A)=M$ to decode all source messages at the destinations, as shown in} Fig. \ref{fig:decoding}. {It is readily concluded that at least $M$ transmissions are required by the relay nodes.
However, if more than $M$ relay transmissions are allowed,
it incurs inefficient spectrum resource usage and the throughput is decreased.
In this sense, we only allow $M$ (the number of source nodes) relay phases over the R-D transmissions in this paper.

Therefore, if $K \ge M$, we can always select the best $M$ relays for transmission.
For instance, method b) can be extended in this case by selecting the best $M$ relays
with the highest achievable CPF rates among all relays in an descending order.
For an extended method c), however, we need to select the best
$M$ vectors
with their highest achievable CPF rates among all relays in an descending order,
with the constraint that the corresponding coefficient vectors linearly
independent.

For the case that $K < M$, by letting some relays to decode more linear combined messages in the first phase (selecting more than one ${\bf a}_m$), $M$ relay
transmission phases are also attainable to make the destinations capable of decoding all source messages.
For example, an implementation algorithm employing an extended method c) is given as follows.}
\begin{itemize}
\item At the end of Phase 1, each relay node tries to decode a number of function messages with
the corresponding achievable CPF rates.
\item Each relay node broadcasts its achievable rates and the message index to a central controller and
it lists different CPF rates in an descending order and selects the highest $M$
CPF rates with their coefficient vectors linearly independent.
\item The central controller notifies each relay its selected rates and the associated message index, followed by
the M relaying transmission phases.
\end{itemize}
{Hence, for the case $K<M$, some relay nodes with higher CPF rates
 may need to transmit in multiple slots.}

In the following, {we seek to find the fundamental limit on the optimal common multicast throughput over the delay-stringent and the delay-tolerant scenarios respectively.}
It is emphasized that the considered relaying phases still consists of $M$ phases.
As discussed above, it follows from the full rank requirement of the constructed matrix
if $K<M$ and more efficient usage of spectrum resources
if $K>M$. Hence, in this work, we only allow $M$ relaying phases.

\section{Delay-Stringent Transmission}
In this section, we shall address the achievable common multicast throughput with CPF under stringent delay constraints, i.e., messages must be delivered within one slot from the source nodes to the destination nodes. In this case, the aim is to maximize the achievable common multicast throughput by optimally allocating the time resources to each phase within each slot. This can be applied to some realtime communication applications with the minimum rate requirement within each slot.

Firstly, we denote $f_{\mathrm{CPF}}$ as the time fraction assigned to the first phase for S-R transmission with CPF and $f_i$ as the time fraction allotted to the $(i+1)$th phase or the $i$th relay for transmission.
In addition, $P_{\mathrm{CPF}}$ and $P_{i}$ are denoted
as the transmit powers of each source node at the first phase
(compute-and-forward phase) and the $i$th relay, respectively.
We also denote {$R_{\mathrm{CPF}}$ and $R_{i}$ as the transmit
 rate of the corresponding phase. Hence, the products
 $f_{\mathrm{CPF}}R_{\mathrm{CPF}}$ and $f_iR_{i}$ are the throughput
 of the CPF phase and the $i$th relay phase, respectively. The end to end
 throughput per slot is hence determined by the minimum of the
 throughput of all phases, namely,
 $\min(f_{\mathrm{CPF}}R_{\mathrm{CPF}},f_iR_{i})$ ($i=1,\cdots,M$).}


Based on the above analysis, the optimization problem within one slot for given channel coefficients of all links, termed as {\bf P1}, is formulated as follows,
\begin{align}
\max_{f_i,f_{\mathrm{CPF}}} \min \bigl( f_{\mathrm{CPF}}R_{\mathrm{CPF}}(P_{\mathrm{CPF}}),f_iR_{i}(P_i) \bigr)  \label{eq:opt_delay}
\end{align}
where $i=1,\ldots M$. The associated power constraint and the physical constraint are given as follows.
\begin{align}
P_{\mathrm{CPF}} \leq P_{0} \label{eq:con1_delay}\\
P_i \leq P_{0} \label{eq:con2_delay} \\
f_{\mathrm{CPF}}+\sum_{i=1}^M f_i \leq 1 \label{eq:con3_delay}
\end{align}
where the objective function in (\ref{eq:opt_delay}) is to maximize
the minimum throughput of each phase.
(\ref{eq:con1_delay}) and (\ref{eq:con2_delay}) give the power
 constraint of the CPF phase and the $i$th relaying phase. Hence, by transmitting at
 $P_0$, each phase achieves its optimal rate.

{Generally speaking, this is a max-min optimization problem and seems
difficult in the first glance.
We hence rewrite {\bf P1} in an equivalent form as follows,}
\begin{align}
\max \quad f_{\mathrm{CPF}}R_{\mathrm{CPF}}(P_0)
\end{align}
subject to
\begin{align}
f_{\mathrm{CPF}}R_{\mathrm{CPF}}(P_0)=f_iR_{i}(P_0)
\end{align}
and the physical constraint in (\ref{eq:con3_delay}).
{This follows from the fact that the achievable rate
is an increasing function of the transmit power and hence
to achieve optimality requires transmission with the highest power allowed.}
This is however a simple linear optimization problem and the solution to {\bf P1} is given by,
\begin{align}
f_{\mathrm{CPF}}^* = \frac{L^*}{R_{\mathrm{CPF}}(P_{sm})} \label{eq:time_delay_relay}\\
f_i^* = \frac{L^*}{R_{i}(P_{r_i})} \label{eq:time_delay_cpf}\\
L^* = \frac{1}{\frac{1}{R_{\mathrm{CPF}}(P_{sm})}+\sum_{i=1}^{M}\frac{1}{R_{i}(P_{r_i})}} \label{eq:throughput_delay}
\end{align}
where the asterisk denotes optimality and $L^*$ is the optimal achievable common multicast throughput with given channel coefficients in delay-stringent applications.

Note that here we consider the simple case that the average power
constraint at each node is imposed as the constraint for each node
for the discussed slot, with the aim to optimize the throughput
within each slot.

\newtheorem{remark}{Remark}
\begin{remark}
A typical application of it is the realtime multicast multimedia which requires the minimum data
rate for the discussed slot. Another typical case would be a very slow fading scenario where
channels can be regarded as quasi-static. For other applications without
the minimum rate constraint or in a relatively fast fading environment, a possible extension
of {\bf P1} is to optimize the throughput within multiple slots by allocating different
power resources to each slot due to channel variations while still maintaining the
delay-stringent constraint, such that the throughput can be improved.
\end{remark}


%

\section{Delay-Tolerant Transmission}
In this section, we shall address the performance achievable by employing compute-and-forward with full channel state information available at the transmitters. We also assume that the derived integer coefficient vectors of all relays are available at all the destination nodes. As it is, all these messages can be obtained via a feedback channel in a time-sharing manner or frequency-sharing manner for each node.

For clarity, we denote $\bar{P}_{\mathrm{CPF}}$ as the average power consumed
at each source node for the CPF phase and $\bar{P}_{i}$ the average power
consumed of the $i$th relay, respectively. Correspondingly,
{$\bar{R}_{\mathrm{CPF}}$ as the average CPF rate achievable
for the CPF phase and $\bar{R}_{i}$ the average rate
at the $i$th relay, respectively.}
Note that all these values are averaged over
the associated channel coefficient distributions. {Hence
$f_i\bar{R}_{\mathrm{CPF}}$ and $f_i\bar{R}_{i}$ are the
average throughput of the CPF phase and the $i$th relaying phase, respectively.
Hence, the throughput of the transmission protocol considered for the delay-tolerant scenario
is given by}
$$\min ( f_{\mathrm{CPF}}\bar{R}_{\mathrm{CPF}}({\bf h}_l), f_i\bar{R}_i({\bf g}_{i})).$$
{To achieve the optimal throughput, we hence need to adjust
power and time resources allocated for each phase.} The associated problem to
maximize the average common multicast throughput, referred to as {\bf P2},
is formulated as follows.


\begin{align}
\max_{R_{\mathrm{CPF}}({\bf h}_l), R_i({\bf g}_i)} \quad \min ( f_{\mathrm{CPF}}\bar{R}_{\mathrm{CPF}}({\bf h}_l), f_i\bar{R}_i({\bf g}_{i})) \label{opt_CSIT}
\end{align}
where $l,i=1,\ldots M$. The associated constraints are given as follows.
\begin{align}
&\bar{P}_{\mathrm{CPF}} \leq P_{0}\label{con1_CSIT}\\
&\bar{P}_i \leq P_{0} \quad i=1,\ldots,M \label{con2_CSIT}\\
&f_{\mathrm{CPF}}+\sum_{i=1}^M f_i \leq 1 \label{con3_CSIT}
\end{align}
The optimization problem above aims to optimally allocate time resources to different phases in order to mitigate the performance degradation caused by bottleneck links. In this sense,
{\bf P2} can also be transformed into an equivalent optimization problem below
\begin{align}
\max_{R_{\mathrm{CPF}}({\bf h}_l), R_i({\bf g}_i)} \quad f_{\mathrm{CPF}}\bar{R}_{\mathrm{CPF}}
\end{align}
subject to the average power constraints in (\ref{con1_CSIT})-(\ref{con2_CSIT}), and the physical constraint in (\ref{con3_CSIT}), and
\begin{align}
&f_{\mathrm{CPF}}\bar{R}_{\mathrm{CPF}}=f_i\bar{R}_i \quad \forall i \label{con4_CSIT}
\end{align}
where (\ref{con4_CSIT}) points out the fact that the product of average rate of each phase and the time resource allotted to that phase should be made equal for each phase for optimality.
Note that it cannot be readily observed that {\bf P2} is a convex optimization problem due to the relationship between $P_{\mathrm{CPF}}$ and $R_{\mathrm{CPF}}$ in (\ref{eqn:app1_2}). Fortunately,
it is verified that $P_{\mathrm{CPF}}$ is indeed a convex function of $R_{\mathrm{CPF}}$ and therefore {\bf P2} is a convex optimization problem. This observation is summarized in Theorem \ref{thm_convex} and the proof is presented in Appendix \ref{appen_1}.
\begin{thm}\label{thm_convex}
{\bf P2} is a convex optimization problem.
\end{thm}
The detailed proof is given in Appendix \ref{appen_1} and omitted here.
According to Theorem \ref{thm_convex},
{\bf P2} can be solved by the Lagrangian multiplier method. The associated Lagrangian multiplier function is given by,
\begin{align}
&F(R_{\mathrm{CPF}},\gamma_i,\beta_i)=f_{\mathrm{CPF}}\bar{R}_{\mathrm{CPF}}
-\beta_0 \left(\bar{P}_{\mathrm{CPF}}-P_{0} \right) \nonumber\\
&- \sum_{i=1}^M \beta_i \left(\bar{P}_{i}-P_{0}  \right) - \sum_{i=1}^M \gamma_i \left( f_{\mathrm{CPF}}\bar{R}_{\mathrm{CPF}}-f_i\bar{R}_{i} \right)\nonumber\\
&-\alpha( f_{\mathrm{CPF}}+\sum_{i=1}^M f_i -1)
\end{align}
where $\beta_0$, $\beta_i$ are the Lagrangian
multipliers with respect to the power constraints of the CPF phase
and the relaying phases. $\gamma_i$ is the Lagrangian multiplier
with respect to the rate constraint. $\alpha$ is the Lagrangian multiplier for the physical
constraint.

The KKT conditions are given by,
\begin{align}
&f_{\mathrm{CPF}}-\beta_0 \cdot \frac{dP_{\mathrm{CPF}}}{dR_{\mathrm{CPF}}} -\sum_{i=1}^L \gamma_i f_{\mathrm{CPF}} = 0 \label{kkt_1}\\
&-2\beta_i\ln 2\frac{2^{2R_{i}}}{g_{r_i}}+\gamma_if_i=0 \label{kkt_2} \\
&\bar{R}_{\mathrm{CPF}} - \sum_{i=1}^M \gamma_i\bar{R}_{\mathrm{CPF}}  - \alpha = 0 \label{kkt_3} \\
&\gamma_i\bar{R}_i-\alpha = 0\label{kkt_4}\\
&f_{\mathrm{CPF}}+\sum_{i=1}^M f_i = 1 \label{kkt_5}
\end{align}

From (\ref{kkt_2}), we then obtain the optimal rate as well as the allotted power for the relaying phases.
\begin{align}
R^*_{r_i} &= \frac{1}{2}\log_2[\frac{f_i^*\gamma_i^*}{2\beta_i^*\ln 2}g_{r_i}]^+, \quad i=1, \cdots, M \label{eqn:rate_relay}\\
P^*_{r_i}&=[\frac{f_i^*\gamma_i^*}{2\beta_i^*\ln 2 }-\frac{1}{g_{r_i}}]^+ \quad i = 1,\cdots, M \label{eqn:power_relay}
\end{align}
where (\ref{eqn:power_relay}) can be directly obtained from the Shannon capacity formula $P=(2^{2R}-1)/g$ and (\ref{eqn:rate_relay}).

For the computation rate, by deriving the first derivative of $R_{\mathrm{CPF}}$ to $P_{\mathrm{CPF}}$ in (\ref{eqn:app1_2}) and insert it into (\ref{kkt_1}) and employing some arithmetic operations, we arrive at the follow equality.
\begin{align}
\frac{l}{\beta_0}=\frac{dx}{(cx-a)^2} \label{equality}
\end{align}
where $l=f_{\mathrm{CPF}}(1-\sum_{i=1}^M \gamma_i)/(2\ln2)$ and $x=2^{2R_{\mathrm{CPF}}}$.
Here we omit the subscripts of $a_m$, $b_m$ and $c_m$ for simplicity and we assume that
$R_{\mathrm{CPF}}^m=\min_i R_{\mathrm{CPF}}$, i.e., the $m$th offers the lowest
CPF rate.
By applying some arithmetic operations on (\ref{equality}), we arrive at a mono basic quadratic equation,
\begin{align}
lc^2x^2-(2acl+\beta_0d)x+a^2l=0
\end{align}
Therefore, the possible solutions to this quadratic equation are given by,
\begin{align}
x=\frac{2acl+ \beta_0 d \pm \sqrt{\beta_0^2d^2+4acld\beta_0} }{2lc^2}
\end{align}

Furthermore, it can be shown that only one out of these two possible solutions to this quadratic equation is feasible.

Firstly, we consider a possible solution, $x_1=\frac{2acl+ \beta_0 d + \sqrt{\beta_0^2d^2+4acld\beta_0} }{2lc^2}$. We then have
\begin{align}
x_1=&\frac{2acl+ \beta_0 d + \sqrt{\beta_0^2d^2+4acld\beta_0} }{2lc^2}\\
&>\frac{2acl}{2lc^2}=\frac{a}{c}
\end{align}
Note that from Lemma \ref{appendix_1}, it is observed that $x<a/c$. Hence this solution is infeasible.

Secondly, for the other possible solution, $x_2=\frac{2acl+ \beta_0 d - \sqrt{\beta_0^2d^2+4acld\beta_0} }{2lc^2}$, we have
\begin{align}
x_2&=\frac{2acl+ \beta_0 d - \sqrt{\beta_0^2d^2+4acld\beta_0} }{2lc^2} \nonumber \\
&<\frac{2acl}{2lc^2}=\frac{a}{c}
\end{align}
On the other hand, since
$$2acl+ \beta_0 d - \sqrt{\beta_0^2d^2+4acld\beta_0}>0$$ we have $x_2>0$.
Hence the feasibility of $x_2$ is demonstrated.
We then arrive at the unique feasible solution to $x=2^{2R_{\mathrm{CPF}}}$, which is,
\begin{align}
x=\frac{2acl+ \beta_0 d - \sqrt{\beta_0^2d^2+4acld\beta_0} }{2lc^2} \label{eqn:kkt_solution}
\end{align}

Therefore, by replacing $x$ with $2^{2R_{\mathrm{CPF}}}$, 
the optimal CPF rate can be obtained accordingly from (\ref{eqn:kkt_solution}).
we hence arrive at Theorem \ref{thm_solution}. 
\begin{thm} \label{thm_solution}
The solution to {\bf P2} is given by,
\begin{align}
R^*_{\mathrm{CPF}}&=\frac{1}{2}\log_2 [\frac{2a_mc_ml+\beta_0^*d_m- \sqrt{(\beta_0^*)^2d_m^2+4a_mc_md_m l \beta_0^*} }{2lc_m^2}]^+ \label{eqn:rate_CDF}\\
R^*_{r_i} &= \frac{1}{2}\log_2[\frac{f_i^*\gamma_i^*}{2\beta_i^*\ln 2}g_{r_i}]^+, \quad i=1, \cdots, M \label{eqn:rate_relay_1}
\end{align}
where we set $l=f_{\mathrm{CPF}}^*(1-\sum_{i=1}^M \gamma_i^*)/(2\ln2)$ and assume that $$R^m_{\mathrm{CPF}}=\min_i R^i_{\mathrm{CPF}}  = R^*_{\mathrm{CPF}}.$$
 Note again that the CPF rate of the $m$th relay is assumed to be the common CPF rate of all relays with the specified channel gains\footnote{The procedure to determine the common CPF rate and the specific index of the associated relay is given at the end of this page.}.

The associated allocated power with respect to the specified channel gains are given as follows,
\begin{align}
P^*_{\mathrm{CPF}}&=[\frac{1-b_m \cdot 2^{2R^*_{\mathrm{CPF}}}}{c_m \cdot 2^{2R^*_{\mathrm{CPF}}}-a_m}]^+\label{eqn:power_CDF}\\
P^*_{r_i}&=[\frac{f_i^*\gamma_i^*}{2\beta_i^*\ln 2 }-\frac{1}{g_{r_i}}]^+ \quad i = 1,\cdots, M \label{eqn:power_relay_1}
\end{align}
\end{thm}

Note that in Theorem \ref{thm_solution} we simply assume the $m$th relay enjoys the minimum CPF rate, i.e., the common CPF rate. However, it may not be so in all cases. A procedure to determine the common CPF rate and the index of the associated relay is therefore given as follows.
\begin{enumerate}
\item Initialization: set $m=1$.
\item Assuming the $m$th relay enjoys the minimum CPF rate, compute $R^m_{\mathrm{CPF}}$ by (\ref{eqn:rate_CDF}) and then compute the CPF power required at all relay nodes, i.e., $P^i_{\mathrm{CPF}}$ ($i=1,\ldots, M$) by (\ref{eqn:power_CDF}) with $R^m_{\mathrm{CPF}}$ as the common CPF rate. If $P^m_{\mathrm{CPF}}=\min_i P^i_{\mathrm{CPF}}$, go to 4) and terminates. Otherwise, go to 3).
\item Set $m=m+1$, go to 2).
\item Output the common CPF rate, the common CPF power required as well as the index of the associated $m$th relay.
\end{enumerate}

According to Theorem \ref{thm_solution} and the algorithm above,
a multi-dimensional bisection search method can be implemented to
obtain the optimal solution to {\bf P2}, with the convergence conditions
in (\ref{kkt_3}) and (\ref{kkt_4}). It is noted that, in each
iteration with the estimated $\beta_i$, $\gamma_i$ ($i=0,\ldots,M$),
one can readily obtain $\bar{R}_i$, $\bar{R}_{\mathrm{CPF}}$ and the
associated average power consumption, we then compute $f_i$ and
$f_{\mathrm{CPF}}$ from (\ref{con4_CSIT}) and (\ref{kkt_5}) and
hence $\alpha$ from (\ref{kkt_4}). After getting all parameters, we
check the sign of the left hand side of (\ref{kkt_3}) and
update $\beta_i$, $\gamma_i$
accordingly. Following this procedure, the optimal solution to {\bf P2}
can be obtained.

{It should be noted that the bisection method adopted can guarantee global convergence
at a very slow convergence rate of $1/2$} \cite{Cheney12}. {However, by carefully selecting the
start intervals of the parameters, the adopted bisection algorithm
is shown to be able to converge to the
global optimal point in tens of iterations in simulation, which is acceptable in
practical implementations.}

\section{Simulation}
We now present some simulation results to compare the achievable rates by employing CPF and DF.
Full channel state information is assumed to be available at the associated
transmitters. The average power constraint at all nodes are assumed to be the same in the simulation setting for simplicity. The channels are assumed to be real valued fading channels and their gains are modeled by zero mean and unit gain Gaussian variables. In addition, the noises at all nodes are additive white Gaussian variables with zero mean and unit variance. For ease of computation, we shall mainly focus on a multicast network with two source nodes, two relay nodes and two destination nodes ($L=M=2$), if not specified.

For comparison, here we briefly give a DF protocol for a multi-source multi-relay multicast network with $2M$ phases, which is,
\begin{itemize}
\item In the first $i$th ($i=1,\ldots,M$) phase, the $i$th source node transmits its data to the $i$th relay nodes at rate $R_i$.
\item In the $(M+i)$th phase ($i=1,\ldots,M$), the $i$th relay node broadcasts the data from the $i$th source node to all destination nodes at rate $R_{M+i}$.
\end{itemize}
For delay-stringent applications, the problem to maximize the common multicast throughput within one slot with DF, is formulated as {\bf P3} below.
\begin{align}
&\max_{f_i} \quad \min_i(f_{i}R_{i}) \label{eqn:DF_DS_opt} \\
s.t.\quad &  \nonumber\\
&P_{i} \leq P_{s_i} \quad i=1,\cdots,M\\
&P_i \leq P_{r_i} \quad i=M+1,\cdots,2M\\
&\sum_{i=1}^{2M} f_i \leq 1
\end{align}
The solution to {\bf P3} is similar to that of {\bf P1} and is hence omitted.
In addition, the average throughput for delay-stringent applications can be readily obtained by averaging over the channel distributions.

Similarly, the problem to optimize the averaged common multicast throughput for delay-tolerant applications is formulated as {\bf P4} below.
\begin{align}
\max_{f_i,R_i({\bf h}_l),R_i({\bf g}_i)} \quad f_{i}\bar{R}_{i} \label{eqn:DF_DT_opt}
\end{align}
subject to the following constraints
\begin{align}
&\bar{P}_{l} \leq P_{s_l}\quad l=1,\cdots,M\\
&\bar{P}_{M+i} \leq P_{r_i} \quad i=1,\cdots,M\\
&f_{l}\bar{R}_{l}=f_i\bar{R}_i \quad \forall l,i \\
&\sum_{i=1}^{2M} f_i \leq 1
\end{align}
Note that {\bf P4} is a standard convex optimization problem and can be readily solved by KKT conditions. The solution to it is however omitted for brevity.

For clarity, a table describing the associated transmit strategies linked to different applications is presented below.
\begin{table}[h!] 
\caption{list of Different Optimization problems with their adopted strategies as well as the related applications}
\centering
\begin{tabular}{|c|c|m{4.7cm}|}
ine
Problem Index &  Strategy   &  Detailed Description  \\
ine
{\bf P1} & CPF-DS & compute-and-forward strategy employed in the delay-stringent application \\
ine
{\bf P2} & CPF-DT & compute-and-forward strategy employed in the delay-tolerant application \\
ine
{\bf P3} & DF-DS & decode-and-forward strategy employed in the delay-stringent application \\
ine
{\bf P4} & DF-DT & decode-and-forward strategy employed in the delay-tolerant application\\
ine
\end{tabular}
\label{Tab1}
\end{table}

Before presenting the performance of the proposed strategies, we would like to show the probability that the network integer vector constructed matrix is not in full rank by employing the proposed methods. It is worth mentioning that, if this matrix is not in full rank, each destination node can not recover all messages from the source nodes. Hence, full rank requirement of the network integer vectors  plays a crucial role in applying CPF strategy.

\begin{figure}[t]
   \centering
   \includegraphics[width = 7.8cm]{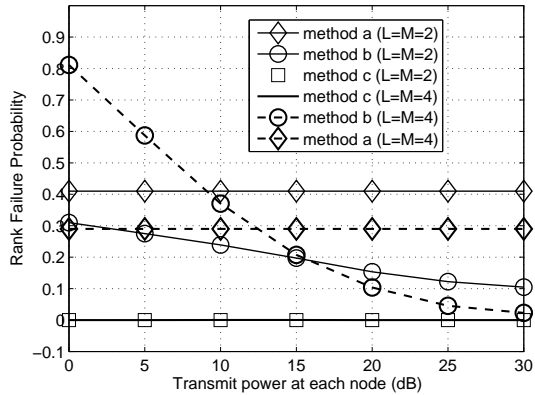}
   \caption{Rank failure probability of CPF employing different methods to
   obtain the integer channel vectors ($K=L=M$).} \label{fig:rank}
   \end{figure}

In Fig.\ \ref{fig:rank}, the probability of rank failure of each method is shown. It is seen that with global optimization, method c) satisfies full rank requirement and the advantage of applying method c) is verified.
Both method a) and method b) have non-zero failure probabilities,
among which method a) has a constant failure probability since its
determination criterion is independent of transmit power. Method b)
is with a decreasing failure probability with the increase of
transmit SNR. Interestingly, with more users at high SNR regime ($K=L=M=4$ at the SNR level over 20dB), it is observed the rank failure probability of method b) is negligible.
On the other hand, it is worth mentioning that using method c) requires additional overhead cost, which would be a critical obstacle for a large-scale multicast network in implementation, as each node needs to exchange some control information on how to construct a global-optimal full network coding system matrix by using CPF.

\begin{figure}[t]
   \centering
   \includegraphics[width = 7.8cm]{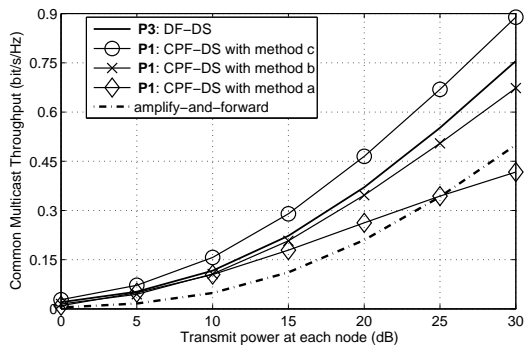}
   \caption{Optimal common throughput for delay-stringent case by
   using different strategies ($K=L=M=2$).} \label{fig:DS_comp}
   \end{figure}

In Fig.\ \ref{fig:DS_comp}, the optimal averaged common multicast throughput by using CPF strategy as well as DF strategy with delay-stringent constraints are compared over $10000$ randomly generated channel realizations.
It is shown that CPF-DS with the integer network channel coefficient vectors found by the global optimal method, i.e., method c), outperforms that with the local optimal method (method b)) and the naive method (method a)). It is also observed that CPF-DS employing method c) outperforms DF in terms of achievable throughput. However, it is seen that CPF-DS with method a) or b) performs worse than DF strategy, due to their non-negligible rank failure probabilities as shown in Fig.\ \ref{fig:rank}.

\begin{figure}[t]
   \centering
   \includegraphics[width = 7.8cm]{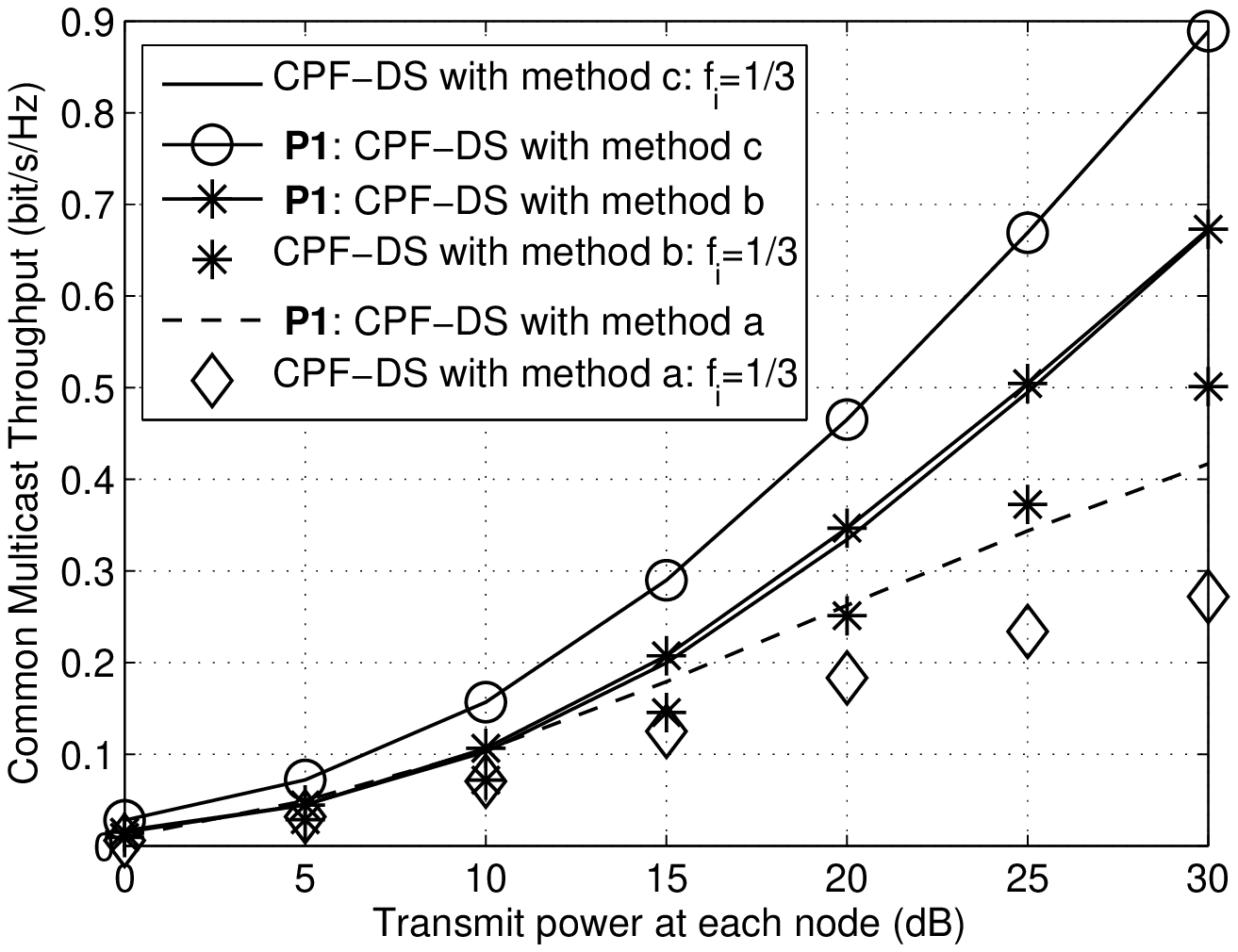}
   \caption{Optimal common throughput for delay-stringent case with
   optimal time allocation and with equal time splitting ($K=L=M=2$). For equal time splitting case, we set $f_{\mathrm{CDF}}=f_1=f_2=1/3$.} \label{fig:CPF_DS_time_comp}
   \end{figure}

In Fig.\ \ref{fig:CPF_DS_time_comp}, the advantage of optimal time allocation is shown for delay-stringent case.
It is observed that the performance of CPF-DS can be greatly improved with optimal time resource allocation. For instance, in the regime of $30$dB transmit power for CPF-DS with method c), an additional $0.2$ bit/s/Hz throughput improvement is achieved  by using optimal time resource allocation, compared with that using equal time-resource allocation.

\begin{figure}[t]
   \centering
   \includegraphics[width = 7.8cm]{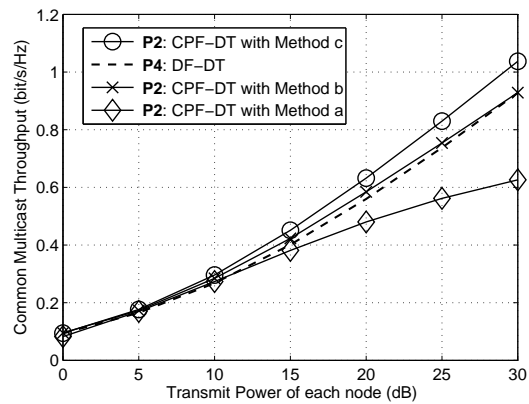}
   \caption{Optimal common throughput for delay-tolerant case by using
   different strategies ($K=L=M=2$).} \label{fig:DT_comp}
   \end{figure}

In Fig.\ \ref{fig:DT_comp}, the optimal common multicast throughput by using different strategies for delay-tolerant applications are shown. It is shown that CPF-DT with method c) outperforms DF strategy in terms of common throughput. For instance, in the regime of 30dB transmit power, employing CPF-DT with method c), over 10\% throughput improvement is achievable, compared with DF-DT. It is also interesting to note that CPF-DT with method b) is slightly better than DF, whereas CPF-DT with method a) is worse than DF,
which is due to the high probability of rank failure and the far-from-optimal
integer coefficient vectors sorted by adopting method a).

\begin{figure}[t]
   \centering
   \includegraphics[width = 7.8cm]{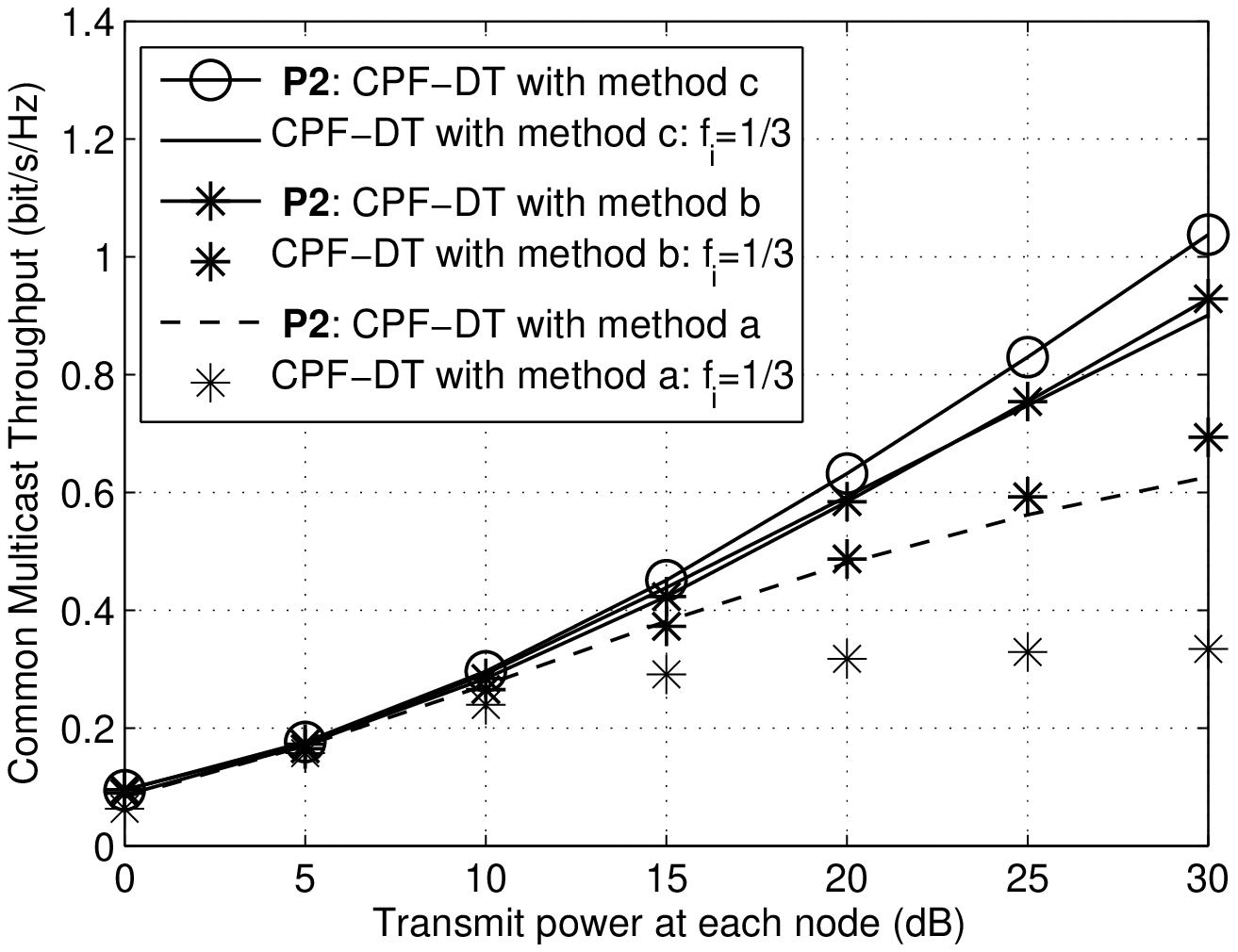}
   \caption{Optimal common throughput for delay-tolerant case
   with optimal time allocation and with equal time splitting ($K=L=M=2$). For equal time splitting case, we set $f_{\mathrm{CDF}}=f_1=f_2=1/3$.} \label{fig:CPF_DT_time_comp}
   \end{figure}

Similar to Fig.\ \ref{fig:CPF_DS_time_comp}, the advantage of optimal time allocation is shown for the delay-tolerant application in Fig.\ \ref{fig:CPF_DT_time_comp}. With transmit power at $30$dB, throughput is increased from roughly $0.9$ bit/s/Hz to over $1$ bit/s/Hz for {\bf P2} with optimal time splitting.

\begin{figure}[t]
   \centering
   \includegraphics[width = 7.8cm]{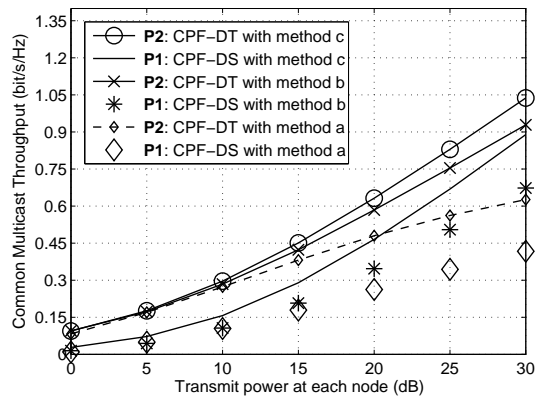}
   \caption{Optimal common throughput for delay-tolerant case
   and delay-stringent case ($K=L=M=2$).} \label{fig:DS_DT}
   \end{figure}

In Fig.\ \ref{fig:DS_DT}, the achievable throughput by using CPF strategy for both the delay-stringent and delay-tolerant cases are compared. It is observed that without delay constraints, higher throughput is expected to be achievable. With transmit power at $30$dB, an additional $0.15$ bit/s/Hz throughput improvement is achieved for CPF-DT compared with CPF-DS, where both of them employ method c).

\begin{figure}[t]
   \centering
   \includegraphics[width = 7.8cm]{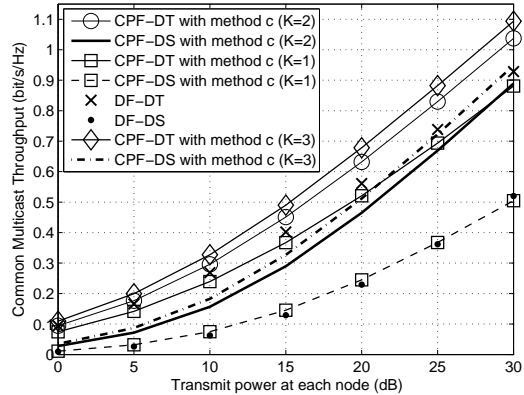}
   \caption{{Optimal common throughput for delay-tolerant case and
    delay-stringent case for two-source, two-destination network. The
    topologies with three relays ($K=3$), two relays ($K=2$) and one relay
    ($K=1$) are evaluated and compared.} } \label{fig:DS_DT_less_relay}
   \end{figure}

In Fig.\ \ref{fig:DS_DT_less_relay},{ we are interested in the topology
with
arbitrary number of relay nodes ($K=3$, $K=2$ and $K=1$). For the case $K=1$,
the sole relay node needs to decode two
function messages for successful source-message decoding at the destination nodes.
It is observed that with less relay nodes, the optimal
achievable common rate with CPF is decreased, for both the delay-stringent and
delay-tolerant scenarios. This is intuitively due to the reduced
cooperative diversity coming from the decreased number of relays. For the case that
$K=3$, it is observed that the optimal throughput of CPF
is further improved than that with $K=2$, which comes from the increased cooperative diversity.

On the other hand,
it is also observed that with the single relay node ($K=1$), the
CPF strategy
performs slightly worse than DF strategy for delay-tolerant case and roughly
as good as DF for delay-stringent case, due to the
reduced relaying diversity.
Taken into account
that more overhead information is required for CPF strategy,
it is intuitively concluded that
DF is still a good choice for transmission in a small-scale network with
less potential relay nodes
than source nodes.}

\begin{figure}[t]
   \centering
   \includegraphics[width = 7.8cm]{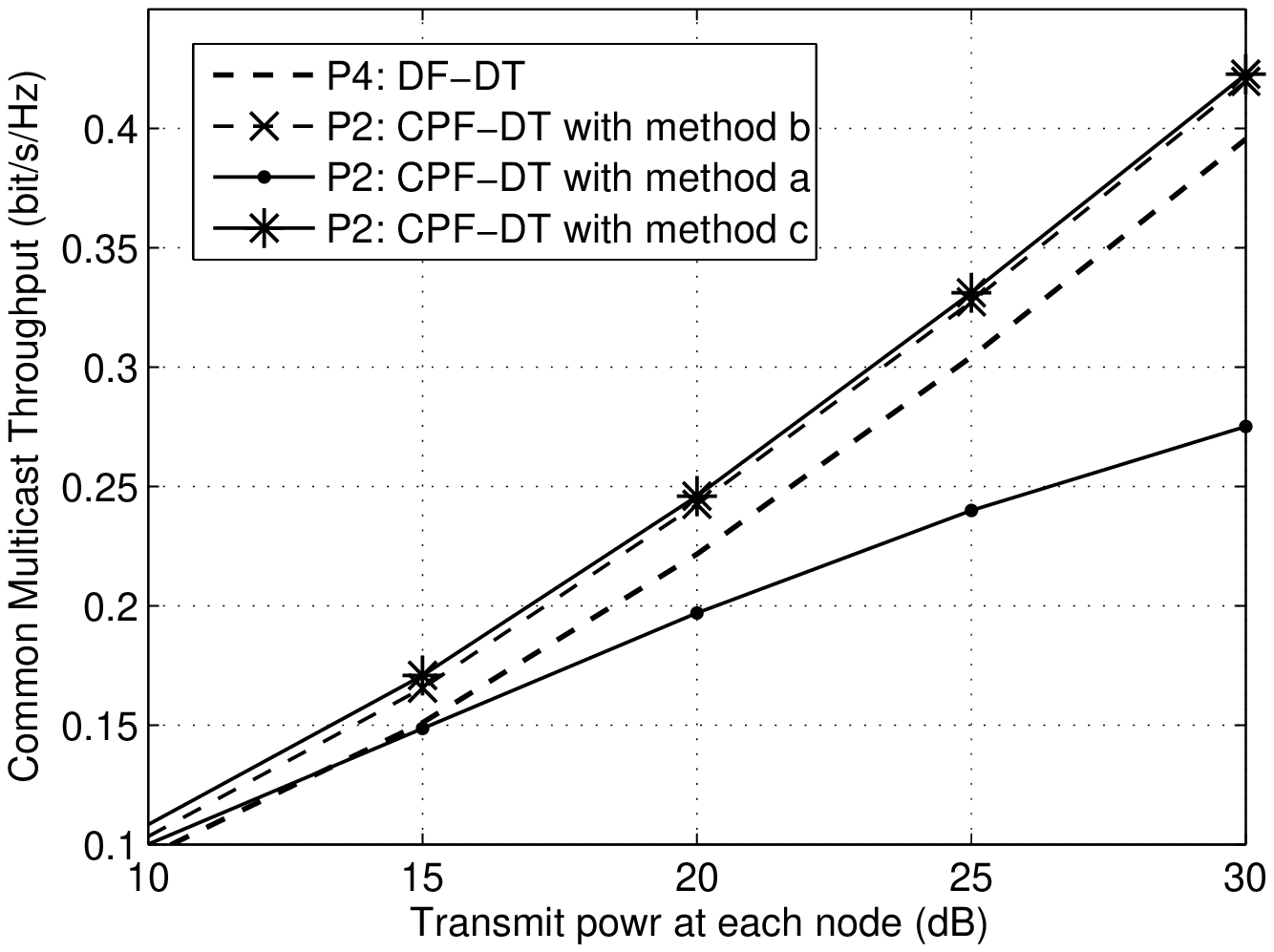}
   \caption{Optimal common throughput for delay-tolerant case
   in a network consisted of four source nodes, four relay nodes
   and four destination nodes ($K=L=M=4$).} \label{fig:four_DT}
   \end{figure}

In Fig.\ \ref{fig:four_DT},
the optimal common multicast throughput for delay-tolerant
application using different strategies is shown for a four-source,
four-relay and four-user multicast network ($K=M=L=4$).
The performance gain of employing CPF over DF is hence verified for
a larger network. It is observed that CPF employing method b) performs
only slightly worse than that with method c) in terms of throughput,
since method b) in a larger network has a lower rank failure probability.
Hence, it is intuitively learned that, it may be worthwhile to employ
method b) for CPF in large networks in the medium to high SNR regime.
In this way, we can not only achieve close to optimal performance as given
by employing method c), but as well avoid the overhead cost incurred
by employing method c).

\section{Conclusion}
In this work, we considered a multi-source multicast network
with the aid of {an arbitrary number of} relay nodes. {We tried to find the fundamental limit on the maximal common
multicast throughput of all S-D pairs}. To this end, a transmission
protocol employing compute-and-forward strategy was proposed {for an arbitrary number of relays}.
Delay-stringent transmission as well as delay-tolerant transmission
applications were both investigated. The associated optimization
problems were formulated and solved, through the allocation of time
and energy resources. {Various simulation was done for validation of the performance improvement of CPF over the conventional DF in terms of throughput.}
{It was shown that with the increasing number of relay nodes, the CPF strategy
can perform better due to the increased diversity.}
Finally, it was intuitively shown that, using CPF with method b) was a good choice
for large networks in medium to high SNR regime, as it not only provided
performance quite close to CPF with method c), but also avoided the
additional communication cost incurred by using method c).

\section*{Acknowledgments}
This work was partially supported by NSFC grant No. 61171064 and The
National 973 Project of China grant No. 2012CB316102.


\appendices

\section{Proof of Lemma \ref{appendix_1}} \label{appen_2}
From (\ref{app1_2}), it is observed that $1-2^{2R_{\mathrm{CPF}}}b_m$ and $2^{2R_{\mathrm{CPF}}}c_m-a_m$ must be both positive or negative, as long as the transmit power is positive. Hence there are two possible scenarios:
\begin{enumerate}
\item $1-2^{2R_{\mathrm{CPF}}}b_m>0$ and $2^{2R_{\mathrm{CPF}}}c_m-a_m>0$,
we then arrive at $\frac{1}{b_m}>2^{R_{\mathrm{CPF}}}>\frac{a_m}{c_m}$.
\item $1-2^{2R_{\mathrm{CPF}}}b_m<0$ and $2^{2R_{\mathrm{CPF}}}c_m-a_m<0$
we then arrive at $\frac{1}{b_m}<2^{2R_{\mathrm{CPF}}}<\frac{a_m}{c_m}$.
\end{enumerate}

Since $a_mb_m-c_m=d_m>0$, we arrive at $\frac{a_m}{c_m}>\frac{1}{b_m}$ and hence only Case 2) is feasible, i.e., $\frac{1}{b_m}<2^{2R_{\mathrm{CPF}}}<\frac{a_m}{c_m}$ holds, as long as a positive transmit power is employed at the relay nodes. Since this property holds for all the relay nodes, Lemma \ref{appendix_1} is proved.

\section{Proof of Convexity of {\bf P2}} \label{appen_1}

Here we shall prove the convexity $P_{\mathrm{CPF}}$ with respect
to $R_{\mathrm{CPF}}$.
Suppose the $m$th relay requires the highest source transmit power
for a common CPF rate. We hence
have $P_{\mathrm{CPF}}=P_{\mathrm{CPF}}^m$.
Therefore, we only need to show $P_{\mathrm{CPF}}^m$ is a convex function of
$R_{\mathrm{CPF}}$.

Recalling that
\begin{align}
R_{\mathrm{CPF}}^m=\frac{1}{2}\log_2^+[\frac{1+P_{\mathrm{CPF}}a_m}{b_m+P_{\mathrm{CPF}}c_m}] \label{app1_1}
\end{align}
and
\begin{align}
P_{\mathrm{CPF}}^m=\frac{1-2^{2R_{\mathrm{CPF}}}b_m}{2^{2R_{\mathrm{CPF}}}c_m-a_m} \label{app1_2_2}
\end{align}
it can be observed that $P_{\mathrm{CPF}}^m$ is a differentiable function of
 $R_{\mathrm{CPF}}$.
Therefore, we only need to show the positivity property of its second derivative with respect
to the associated CPF rate, $R_{\mathrm{CPF}}$.
To proceed, the first derivative with respect to $R_{\mathrm{CPF}}$ is given by,
\begin{align}
\frac{dP_{\mathrm{CPF}}^m}{dR_{\mathrm{CPF}}}&=-\frac{1}{\log_2 e}\frac{2b \cdot 2^{2R_{\mathrm{CPF}}}}{2^{2R_{\mathrm{CPF}}}c_m-a_m} \nonumber \\
&-\frac{2^{2R_{\mathrm{CPF}}} \cdot 2c_m}{\log_2 e}\frac{1-2^{2R_{\mathrm{CPF}}}b_m}{(2^{2R_{\mathrm{CPF}}}c_m-a_m)^2} \label{app1_3}\\
&=\frac{2}{\log_2 e} \frac{2^{2R_{\mathrm{CPF}}}(a_mb_m-c_m)}{(2^{2R_{\mathrm{CPF}}}c_m-a_m)^2} \label{app1_4}\\
&=\frac{2}{\log_2 e} \frac{2^{2R_{\mathrm{CPF}}}d_m}{(2^{2R_{\mathrm{CPF}}}c_m-a_m)^2}>0 \label{derivative}
\end{align}
which confirms the fact that $P_{\mathrm{CPF}}^m$ is an increasing function of $R_{\mathrm{CPF}}$.

The second derivative is derived as follows similarly,
\begin{align}
&\frac{d^2P_{\mathrm{CPF}}^m}{dR_{\mathrm{CPF}}^2} \nonumber\\
=&\frac{4}{\log_2^2 e} \left( \frac{d_m2^{2R_{\mathrm{CPF}}}}{(2^{2R_{\mathrm{CPF}}}c_m-a_m)^2}-\frac{2c_md_m2^{4R_{\mathrm{CPF}}}}{(2^{2R_{\mathrm{CPF}}}c_m-a_m)^3} \right) \label{app1_6}\\
=&\frac{4}{\log_2^2 e}\frac{2^{2R_{\mathrm{CPF}}}d_m(2^{2R_{\mathrm{CPF}}}c_m-a_m)-2c_md_m2^{4R_{\mathrm{CPF}}}}{(2^{2R_{\mathrm{CPF}}}c_m-a_m)^3}\label{app1_7}\\
=&-\frac{4}{\log_2^2 e}\frac{a_md_m2^{2R_{\mathrm{CPF}}}+c_md_m2^{4R_{\mathrm{CPF}}}}{(2^{2R_{\mathrm{CPF}}}c_m-a_m)^3} \label{app1_8}\\
>&0 \label{app1_9}
\end{align}
where (\ref{app1_9}) comes from the fact that
$(2^{2R_{\mathrm{CPF}}}c_m-a_m)^3<0$ in Lemma \ref{appendix_1} and
the negative sign in (\ref{app1_8}). Hence $P_{\mathrm{CPF}}^m$ is a
convex function of $R_{\mathrm{CPF}}$.

\end{document}